# Characterization of crystal structure and precipitation crystallography of a new $Mg_xAl_{2-x}Gd$ phase in $Mg_{97}Al_1Gd_2$ alloy

**X.-F. Gu**, and **T. Furuhara**

*Institute for Materials Research, Tohoku University, Sendai, 980-8577, Japan*

## ABSTRACT

The composition, crystal structure and precipitation crystallography of a newly found precipitate are characterized by a Cs-corrected scanning transmission electron microscopy (STEM). The composition of the plate-like precipitate could be expressed as $Mg_xAl_{2-x}Gd$ (x = 0.38), and its crystal structure is the same as the fcc type Laves phases $Mg_2Gd$ and $Al_2Gd$ with the lattice parameter of 7.92 Å (space group No. 227, Fd-3m). The orientation relationship between the matrix and precipitate is found to be $(0001)_m // (111)_p$ and $[10\bar{1}0]_m // [1\bar{1}0]_p$, and the habit plane is parallel to $(0001)_m // (111)_p$ plane. In addition, this preferred crystallography of phase transformation is well explained based on the atomic matching in the interface.

**Introduction**

Mg alloy containing rare earth element (such as Gd) has shown prominent precipitation hardening and promising property at high temperature (Apps *et al.*, 2003; Nishijima *et al.*, 2006; Nie *et al.*, 2008; Vostrý *et al.*, 1999). The precipitation sequence in Mg-Gd alloy is found to be as follows: supersaturated solution α-Mg → β"(DO19) → β'(bco) → $β_1$(fcc) → β(fcc) (Apps *et al.*, 2003; Gao *et al.*, 2006). The addition of other alloy elements, such as Ag or Zn, is aimed to enhance the response of precipitation hardness and to reduce the amount of rare earth element Gd (Yamada *et al.*, 2009; Gao & Nie, 2008). Especially for Zn added Mg-Gd alloys, different precipitation sequence has been found, and GP zone (Saito *et al.*, 2011; Nishijima *et al.*, 2008; Nie *et al.*, 2008) and a long-period stacking ordered (LPSO) structure can form during ageing (Yamasaki *et al.*, 2007). Similarly, Kishida et al. (Kishida *et al.*, 2013; Kishida *et al.*, 2012; Yokobayashi *et al.*, 2011) shows that the addition of Al to Mg-Gd alloy will also help to form a well ordered LPSO structure in high Mg-Al-Gd alloys, such as in Mg-3.5at.%Al-5.0at.%Gd alloy. In addition, reproducible precipitation crystallography, such as orientation relationship and interfacial orientation, has been found in those alloys (Nishijima *et al.*, 2006; Gao *et al.*, 2006; Yokobayashi *et al.*, 2011; Apps *et al.*, 2003), and the orientation relationship and interface between product phase and matrix is summarized in Table 1. In contrast to observed phases shown in Table 1, we will report a new phase found during annealing of Mg-Al-Gd alloy. The composition, crystal structure and its precipitation crystallography will be characterized.

Experimental characterization of atomic structure has been facilitated with advancement of electron microscopy, especially with atomic-resolution scanning transmission electron microscopy (STEM). Relative heavy atoms show brighter contrast than that for light atoms in high angle annular dark field (HAADF)-STEM image, which is so called Z (atomic-number) contrast image. Therefore, HAADF-STEM is particularly useful to reveal the segregation and/or atomic structure of the precipitates with large atomic-number difference, such as in magnesium alloys containing rare earth elements (Kishida *et al.*, 2013; Kishida *et al.*, 2012; Yokobayashi *et al.*, 2011; Egusa & Abe, 2013; Zhu et al., 2016; Zhu et al., 2010; Kishida et al., 2015; Nie et al., 2016). With the advent of Cs-corrected STEM, the STEM resolution is remarkably improved to sub-angstrom level. As a result, the observed atomic structure can be directly compared with the proposed crystal structure, and this method will be applied in this work.

**Experimental method**

A $Mg_{97}Al_1Gd_2$ (at.%, nominal composition) alloy was prepared by using induction melting under the protection of Argon atmosphere. Then the casted sample is isothermally held at the temperatures between 500°C and 600°C for 2h originally designed for solution treatment. After the heat treatment, the sample is quenched in water. The microstructure is characterized by SEM (JEOL 7001F), Electron Backscatter Diffraction (EBSD, TSL) technique and Cs-corrected HAADF-STEM (Titan$^{3TM}$ G2 60-300, 300 kV; FEI). The convergent semi-angle used for HAADF-STEM observation was 21.4 mrad. The collection semi-angle of angular detector for HAADF-STEM images was 50.5-200 mrad. TEM specimens of ϕ3 mm with a thickness of 0.5 mm were cut from the heat treated specimens by using electro-discharge machining. The discs were then grinded to about 60 μm and thinned by dimpling grinder (Model 2000, Fischione) followed by low-energy ion milling from 4 kV to 500 V (PIPS model691; Gatan Inc.) with an incidence angle of 4°. Chemical composition is measured by SEM-EDS (Energy Dispersive X-ray Spectroscopy, EDAX Inc.) or (S)TEM-EDS (Bruker, Inc.). The STEM images in this study are simulated with COMPUTEM (Kirkland, 2010), and the atomic structures and diffraction patterns are simulated with PTCLab (Gu *et al.*, 2016).

**Results**

Firstly, the microstructure of as-casted and aged sample observed by SEM will be examined. Figure 1a shows the microstructure of as-casted sample, and the element mappings of Gd and Al are shown in 1b and 1c, respectively. According to the element content, the particle in Figure 1a is found to be $Al_2Gd$ as indicated by single arrow, which is formed as part of eutectic phase along the grain boundaries. In addition, it is also found Gd enriches at grain boundary. Figure 1d shows the microstructure of the sample aged at 520°C for 2h. Parallel precipitates are formed within each grain due to the supersaturated solute atoms during casting, and diffusion of the solute atoms from the segregated grain boundaries. There are remained $Al_2Gd$ particles after annealing, which may due to higher melting point of the Laves phase $Al_2Gd$ (1798K) and thus strong binding between Al and Gd atoms. The orientation of hcp matrix is overlaid on each grain according to the EBSD results. It is noted that all of the interface traces are normal to $[0001]_m$ direction in matrix, thus the habit plane (HP) of the precipitate should be parallel to $(0001)_m$ plane, which is the same as LPSO phase in Table 1 but different from other precipitates in the table. Next, the composition and structure of the precipitate will be determined.

Since the thickness of the precipitate is too small to measure the composition with SEM-EDS, TEM-EDS is applied. The composition is shown in Table 2. Accordingly, the summation of atomic percentage of Al and Mg is about two times of Gd content, which is totally different from the reported compositions of LPSO phases with Mg content larger than 75 at.% (Kishida *et al.*, 2015) and other binary compounds in Table 1. Therefore, the observed precipitate is a new phase in Al added Mg-Gd alloy. It is noted that the elements Al and Gd are easy to form fcc type Laves phase of $Al_2Gd$, while Mg and Gd tend to form Laves phase of $Mg_2Gd$ with the same structure as $Al_2Gd$. Therefore, it is highly possible that the found precipitates also have the same structure as $Mg_2Gd$ and $Al_2Gd$, which is shown in Figure 2a, and its composition could be expressed as $Mg_xAl_{2-x}Gd$ where x = 0.38 in present study. In the structural model shown in Figure 2a, Gd atoms are located at the internal coordinate (0,0,0) in the coordinate system of crystal, while Mg/Al atoms are located at the internal coordinate of (0.625, 0.625, 0.625), and among the Mg/Al atoms, 19% of the atoms are magnesium elements. In order to confirm the proposed crystal structure in Figure 2a, the atomic positions are directly compared to HAADF-STEM images taken by a Cs-corrected HAADF-STEM. The result is shown in Figure 2b-c from two zone axes with 30° difference, i.e. $[1\bar{2}1]_p$ and $[1\bar{1}0]_p$ where subscript "p" denotes the plane or direction in precipitates. The bright dumbbells in the images are Gd atoms according to the Z contrast in HAADF-STEM images, while the weak contrast in the image should correspond to Mg or Al atoms. The simulated atomic structures are superimposed on both HAADF-STEM images, and they agree well with the observations. In the simulation, the lattice parameter of 7.92 Å is used which is close to that of $Al_2Gd$ 7.902 Å (Baenziger & Moriarty, 1961), and this lattice parameter also agrees with the diffraction pattern as will be shown latter.

In Figure 1d, the precipitates exhibit reproducible morphology, and it is probably that there is a unique orientation relationship between matrix and precipitates. Figures 3 and 4 show the atomic structures around the edge-on interface and their corresponding diffraction patterns from two different zone axes similar to Figure 2. The diffraction pattern of precipitate could be indexed with the crystal model in Figure 2a, and the simulated diffraction of matrix, precipitate and their combinations are shown in (c), (d) and (e) in Figures 3 and 4, respectively. The flat interface is parallel to $(0001)_m$ consistent with the observation in Figure 1d. It is also noted that the growth of precipitate is proceeded by a ledge mechanism parallel to $(0001)_m$, as indicated by double arrows in Figure 3a and 4a. According to diffraction pattern or atomic structure in these figures, the OR could be expressed as:

$$(0001)_m // (111)_p$$

$$[10\bar{1}0]_m // [1\bar{1}0]_p \qquad (1)$$

$$[\bar{2}110]_m // [11\bar{2}]_p$$

with the major interface or called habit plane paralleling to the base plane in matrix,

$$(0001)_m // (111)_p. \qquad (2)$$

This preferred crystallography usually can be understood by examining the atomic matching in the interface (Zhang & Kelly, 2009; Zhang & Weatherly, 2005), where good matching in the interface means lower structure part of interfacial energy, thus the good matching interface will be formed during precipitation. In this work, we examine the atomic matching at the interface for all possible interfacial orientations at the measured orientation relationship by considering the percentage of good matching sites among the atoms on each interface (Furuhara *et al.*, 2006; Furuhara *et al.*, 2002; Yang & Zhang, 2012), where a good matching site is defined as the atomic distance between an atomic pair from two crystals smaller than a certain criterion $15\%|a_m|$ where $a_m$ is the lattice parameter of matrix in base plane (Liang & Reynolds, 1998). The calculated result is shown in Figure 5. In Figure 5a, the percentage of good matching site in all possible interfaces in three dimension space is shown in a $[0\bar{1}10]_m$ stereographic projection, and the best matching interface is parallel to $(0001)_m$, which is in good agreement with observed habit plane. The second best matching plane is $\{\bar{2}110\}_p$ which could be the side facets of the precipitate. The detailed atomic matching in the $(0001)_m$ interface is shown in Figure 5b. According to the lattice parameters of product phase and matrix, the misfit in the interface is isotropic and around 0.7%. All the lattices of precipitate nearly coincide with the matrix and form a near coincidence site lattice (CSL). True CSL could be formed by an isotropic constrained strain about 0.7%. The unit cell of the CSL is three times of that in matrix and the interface could be called $\Sigma 3$ interface. Therefore, it is clear that the formation of $(0001)_m$ will result in a low energy interface due to good matching in this plane. The preference of the orientation relationship in Equation (1) is possibly due to the formation of this well matched interface.

## Summary

In this work, a new precipitate is characterized by a Cs-corrected STEM. The composition of this plate-like precipitate is $Mg_xAl_{2-x}Gd$ (x = 0.38) with similar crystal structure as fcc type Laves phase $Mg_2Gd$ or $Al_2Gd$ with a lattice parameter of 7.92 Å. The orientation relationship between Mg matrix and the precipitate is found to be $(0001)_m$ // $(111)_p$ and $[10\bar{1}0]_m$ // $[1\bar{1}0]_p$ with the habit plane paralleling to $(0001)_m$ // $(111)_p$. According to the atomic matching at the interface, the observed habit plane is the best matching interface, and could be considered as a low energy Σ3 boundary with a negligible constrain.

## Acknowledgement

This work was supported by a Grant-in-Aid for Scientific Research on Innovative Areas, "Synchronized Long-Period Stacking Ordered Structure", from the Ministry of Education, Culture, Sports, Science and Technology, Japan (No.23109001). The Mg alloy samples used in the present study were supplied by Kumamoto University. Special thanks to technical support of TEM by Mr. Y. Hayasaka from the Electron Microscopy Center, Tohoku University.

## Reference

Apps, P. J., Karimzadeh, H., King, J. F. & Lorimer, G. W. (2003). *Scripta Mater.* **48**, 1023-1028.
Baenziger, N. C. & Moriarty, J. L. (1961). *Acta Crystallogr.* **14**, 948-950.
Egusa, D. & Abe, E. (2013). *Acta Mater.* **60**, 166-178.
Furuhara, T., Maki, T. & Kimori, T. (2006). *Metall. Mater.Trans. A* **37**, 951-959.
Furuhara, T., Oishi, K. & Maki, T. (2002). *Metall. Mater.Trans.* **33A**, 2327-2335.
Gao, X., He, S. M., Zeng, X. Q., Peng, L. M., Ding, W. J. & Nie, J. F. (2006). *Mater. Sci. Eng., A* **431**, 322-327.
Gao, X. & Nie, J. F. (2008). *Scripta Mater.* **58**, 619-622.
Gu, X.-F., Furuhara, T. & Zhang, W.-Z. (2016). *J. Appl. Cryst.*, **49**, doi:10.1107/S1600576716006075.
Kirkland, E. J. (2010). *Advanced computing in electron microscopy*. Springer Science & Business Media.
Kishida, K., Nagai, K., Matsumoto, A., Yasuhara, A. & Inui, H. (2015). *Acta Mater.* **99**, 228-239.
Kishida, K., Yokobayashi, H. & Inui, H. (2013). *Philos. Mag.* **93**, 2826-2846.
Kishida, K., Yokobayashi, H., Inui, H., Yamasaki, M. & Kawamura, Y. (2012). *Intermetallics* **31**, 55-64.
Liang, Q. & Reynolds, W. T., Jr. (1998). *Metall. Mater.Trans.* **29A**, 2059-2072.
Nie, J. F., Oh-ishi, K., Gao, X. & Hono, K. (2008). *Acta Mater.* **56**, 6061-6076.
Nie, J. F., Wilson, N. C., Zhu, Y. M. & Xu, Z. (2016). Acta Mater. **106**, 260-271.
Nishijima, M., Hiraga, K., Yamasaki, M. & Kawamura, Y. (2006). *Mater. Trans.* **47**, 2109-2112.
Nishijima, M., Hiraga, K., Yamasaki, M. & Kawamura, Y. (2008). *Mater. Trans.* **49**, 227-229.

Saito, K., Yasuhara, A. & Hiraga, K. (2011). *J. Alloys Compd.* **509**, 2031-2038.
Vostrý, P., Smola, B., Stulikova, I., Von Buch, F. & Mordike, B. L. (1999). *Phys. Status Solidi.* **175**, 491-500.
Yamada, K., Hoshikawa, H., Maki, S., Ozaki, T., Kuroki, Y., Kamado, S. & Kojima, Y. (2009). *Scripta Mater.* **61**, 636-639.
Yamasaki, M., Sasaki, M., Nishijima, M., Hiraga, K. & Kawamura, Y. (2007). *Acta Mater.* **55**, 6798-6805.
Yang, X.-P. & Zhang, W.-Z. (2012). *Sci. China Technol. Sc.* **55**, 1343-1352.
Yokobayashi, H., Kishida, K., Inui, H., Yamasaki, M. & Kawamura, Y. (2011). *Acta Mater.* **59**, 7287-7299.
Zhang, M. X. & Kelly, P. M. (2009). *Prog. Mater. Sci.* **54**, 1101-1170.
Zhang, W. Z. & Weatherly, G. C. (2005). *Prog. Mater. Sci.* **50**, 181-292.
Zhu, Y. M., Morton, A. J. & Nie, J. F. (2010). Acta Mater. **58**, 2936-2947.
Zhu, Y. M., Oh-ishi, K., Wilson, N. C., Hono, K., Morton, A. J. & Nie, J. F. (2016). Metall. Mater.Trans. A **47**, 927-940.

**Tables**

Table 1 Orientation relationship (OR) found in Mg-Gd alloys

|  | Phase | Morphology | OR | Interface | Ref. |
|---|---|---|---|---|---|
| **Mg-Gd** | β" (DO19) | Plate or rod | $[0001]_p//[0001]_m$ $(10\bar{1}0)_p//(10\bar{1}0)_m$ | $(2\bar{1}\bar{1}0)_m$ | (Apps *et al.*, 2003) |
|  | β' (bco) | Globular | $[001]_p//[0001]_m$ $(010)_p//(\bar{1}100)_m$ | irregular | (Nishijima *et al.*, 2006) |
|  | $β_1$ (fcc) | Plate | $[110]_p//[0001]_m$ $(\bar{1}12)_p//(1\bar{1}00)_m$ | $(1\bar{1}00)_m$ | (Gao *et al.*, 2006) |
|  | β (fcc) | Plate | $[110]_p//[0001]_m$ $(\bar{1}12)_p//(1\bar{1}00)_m$ | $(10\bar{1}0)_m$ | (Gao *et al.*, 2006) |
| **Mg-Al-Gd** | LPSO | plate | $[1\bar{1}00]_p//[2\bar{1}\bar{1}0]_m$ $(0001)_p//(0001)_m$ | $(0001)_m$ | (Yokobayashi *et al.*, 2011) |

Table 2 Composition of lamellar precipitate measured by STEM-EDS

| Element | wt.% | at.% |
|---|---|---|
| **Mg** | 9.0 | $25.7 \pm 0.2$ |
| **Al** | 16.0 | $41.2 \pm 0.2$ |
| **Gd** | 75.0 | $33.1 \pm 0.35$ |

**Figures**

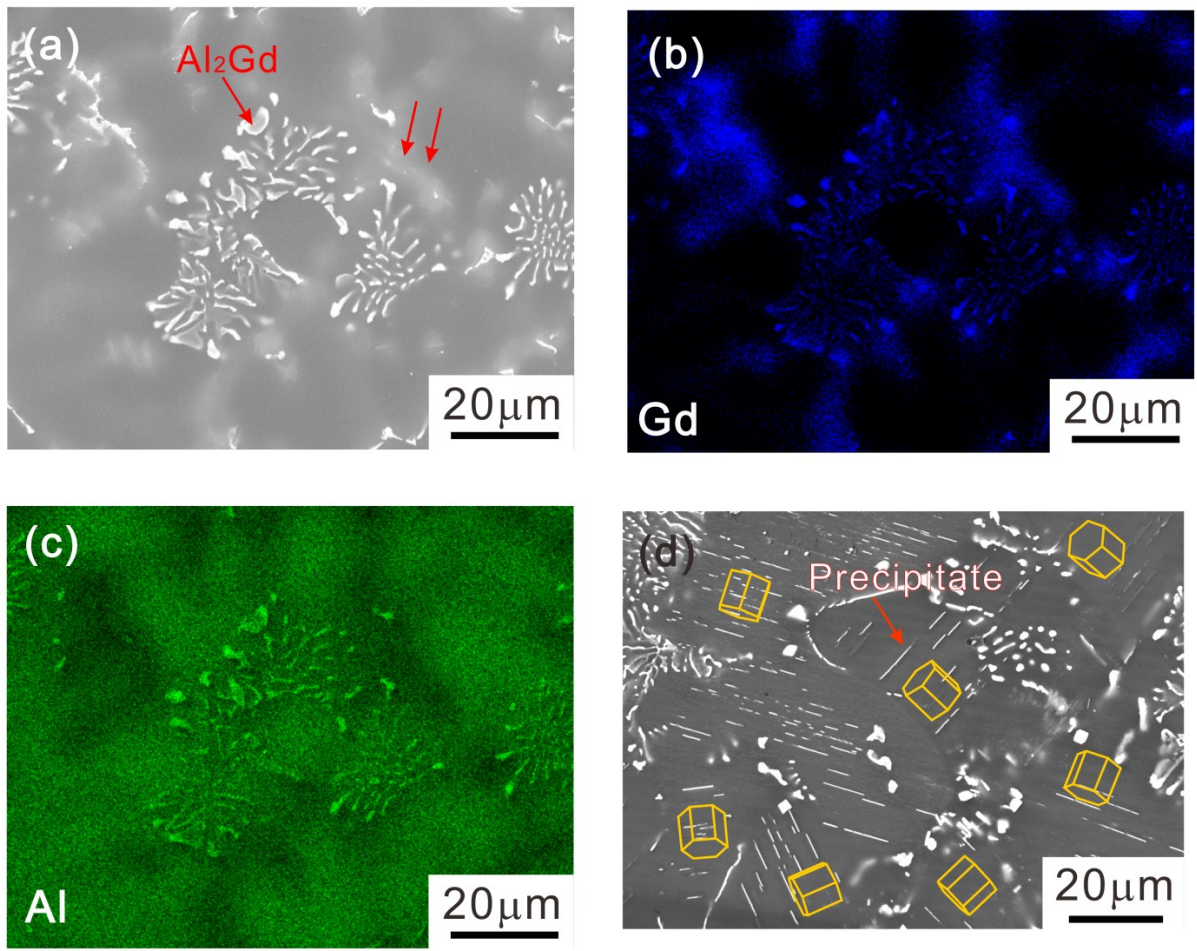

Figure 1 Microstructure of as-casted and aged sample observed by SEM. (a) As-casted sample with $Al_2Gd$ particle (single arrow) and Gd segregation (double arrows), (b) Gd element mapping in (a), (c) Al element mapping in (a), (d) Plate-like precipitates during isothermal holding at 520°C for 2h.

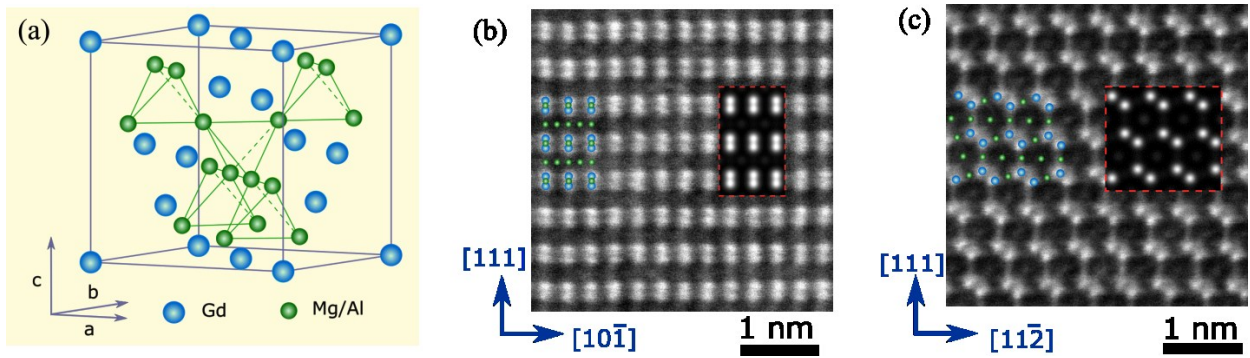

Figure 2 Structure model for the precipitate (a), and atomic resolution HAADF-STEM images (raw data) of product phase at (b) $[1\bar{2}1]_p$ zone axis , (c) $[1\bar{1}0]_p$ zone axis (this sample may deviate a little from the zone axis). The inserted image within the dashed frame is simulated images, and the rigid atomic structures are also overlaid on the images. The simulation parameters are performed with an accelerating voltage of 300 kV, convergent semi-angle of 21.4 mrad, the annular detector angle range of 50.5-200 mrad. In addition, the pixels size of 256 by 256 is used, and the sample along the beam axis is divided with a slice thickness of 0.2 nm.

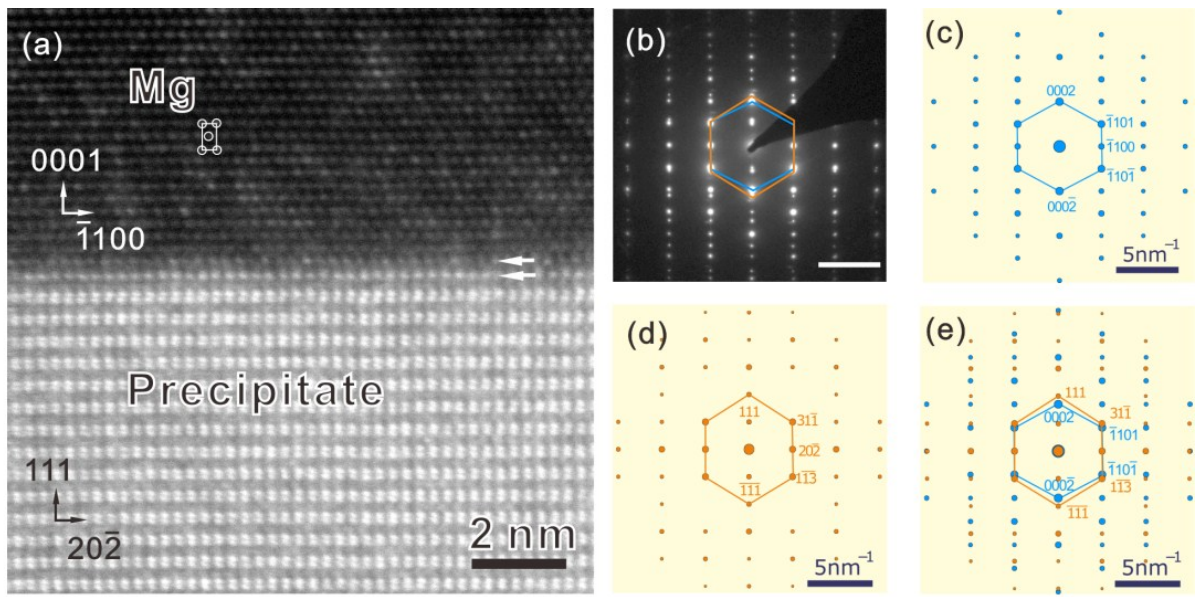

Figure 3 (a) Atomic resolution HAADF-STEM images (raw data) of the edge-on interface between Mg matrix and product phase at $[11\bar{2}0]_m$ // $[1\bar{2}1]_p$ zone axis, (b) Corresponding diffraction pattern, (c) Simulated diffraction pattern of Mg matrix in (b), (d) Simulated diffraction pattern of product phase in (b), (e) Superimposed diffraction pattern of (c) and (d).

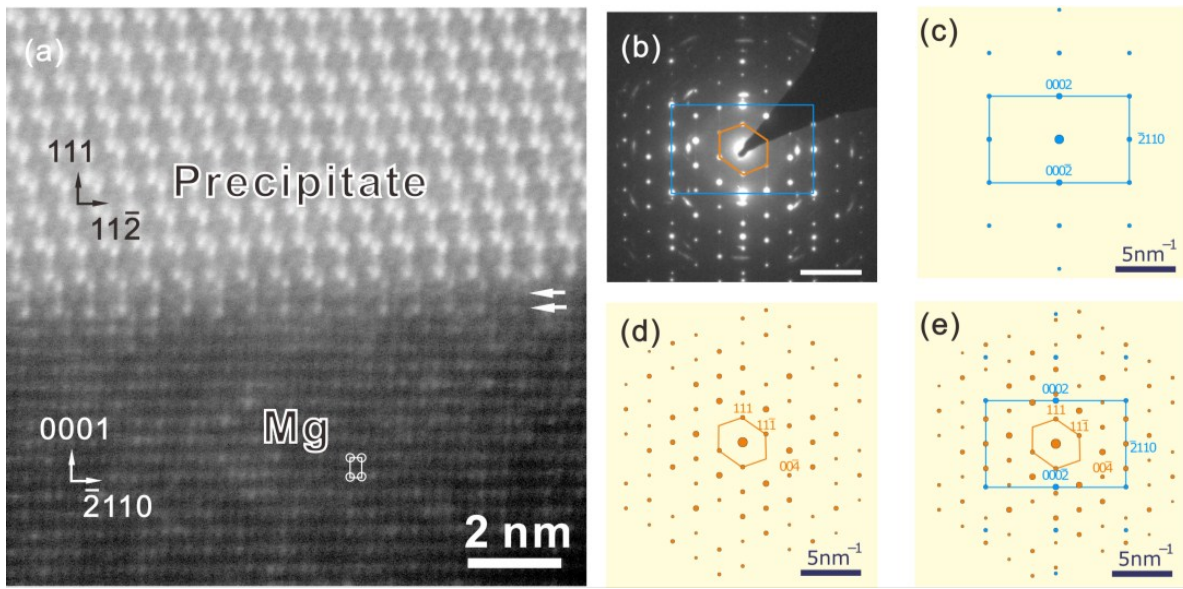

Figure 4 (a) Atomic resolution HAADF-STEM images (raw data) of the edge-on interface between Mg matrix and product phase at $[10\bar{1}0]_m // [1\bar{1}0]_p$ zone axis, (b) Corresponding diffraction pattern, (c) Simulated diffraction pattern of Mg matrix in (b), (d) Simulated diffraction pattern of product phase in (b), (e) Superimposed diffraction pattern of (c) and (d).

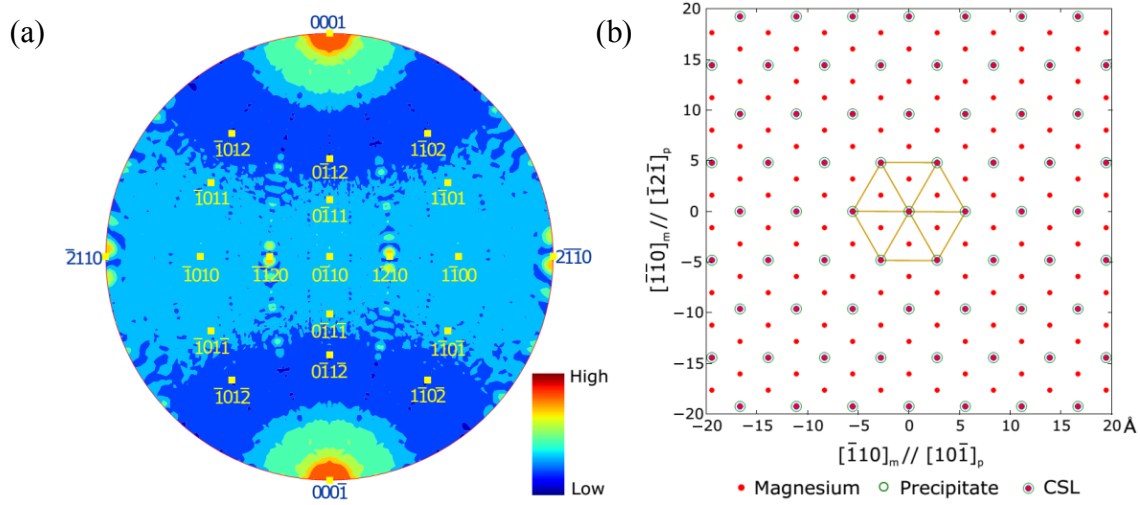

Figure 5 Atomic matching between the precipitate and Mg matrix at the measured OR. (a) Fraction of good matching site on all possible interfaces in a $[0\bar{1}10]_m$ stereographic projection, (b) Atomic matching at $(0001)_m // (111)_p$ interface.